\documentclass[showpacs,preprint,aps]{revtex4}
\usepackage{graphicx}
\usepackage{dcolumn}
\usepackage{bm}
\begin{document}
\setcounter{page}{1}
\title
{Analysis of averaged multichannel delay times}
\author
{N. G. Kelkar and M. Nowakowski}
\affiliation{ Departamento de Fisica, Universidad de los Andes,
Cra.1E No.18A-10, Santafe de Bogota, Colombia}
\begin{abstract}
The physical significances and the pros and cons involved in the usage 
of different time delay formalisms are discussed. 
The delay time matrix introduced by Eisenbud, where only $s$-waves 
participate in a reaction, is in general related to the definition of 
an angular time delay which is shown not to be equivalent to the so-called 
phase time delay of Eisenbud and Wigner even for single channel scattering. 
Whereas the expression due to Smith which is 
derived from a time delayed radial wave packet is 
consistent with a lifetime matrix which is Hermitian, this is not true
for any Eisenbud-type lifetime matrix which violates time reversal 
invariance. 
Extending the angular time delay of Nussenzveig to multiple channels, 
we show that if one performs an average over the directions and subtracts 
the forward angle contribution containing an interference of the 
incident and scattered waves, the 
multichannel angle dependent average time delay reduces to the one 
given by Smith. The present work also rectifies a recently  
misinterpreted misnomer of the relation due to Smith. 
\end{abstract}
\pacs{03.65.Nk, 11.80.Gw}
\maketitle
\section{Historical development of different delay time concepts} 
The concept of time delay was introduced around the year 1950 by 
different authors \cite{eisen,wigner,others}. 
Eisenbud and Wigner introduced it \cite{eisen, wigner} 
by following the peak of spherical wave packets. 
This approach is widely used in 
tunneling and is commonly known as a phase time delay since it involves the 
energy derivative of the scattering phase shift. 
Eisenbud \cite{eisen} in his thesis which remained unpublished, also considered 
multichannel time delay, however, in the context of 
$s$-wave scattering only. He obtained an expression which is closely 
related to the angular time delay as defined by Froissart, Goldberger and
Watson \cite{goldb}. An entirely different approach based on the concept of
``dwell time" was introduced by F. T. Smith in 1960 \cite{smith}. The 
average dwell time had a rather straightforward meaning of the average time
spent by particles interacting in a certain region (with the average 
being over the scattering channels). The time delay was then the difference 
of the time spent with and without interaction. 
This concept became quite popular in the
later years in different branches of physics 
and is also used currently in the context of chaotic 
scattering \cite{chaos}. The average dwell time concept finds application in 
tunneling problems \cite{haugewin} too, where the average is over the 
reflection and transmission channels. Smith defined a lifetime matrix, 
${\bf Q}$, which is Hermitian and is nicely related to the 
scattering ${\bf S}$ matrix, such 
that one can even define a time operator and find a 
relation which when inverted allows one to compute ${\bf S}$ from ${\bf Q}$. 
Indeed, B. A. Lippmann defined a similar 
time delay operator in 1966 \cite{lipp}. 
A mathematically rigorous 
discussion of the multichannel time delay as derived by Smith
can be found in an article by Ph. A. Martin \cite{martin}. 
The application of delay time relations to study resonances 
ranges from atomic physics \cite{atomic} to 
particle and nuclear physics, where, the method 
was applied to study baryon, meson and pentaquark unstable states 
\cite{me1,usall}.      

\section{Time delayed wave packets} 
The discussions on delay times often begin with a narrow one-dimensional 
wave packet, where one tries to answer at what value of $x$ would the 
wave packet $\Psi(x,t)$ be peaked at a given point of time $t$. 
It is in this spirit that Wigner started with a time dependent wave function 
composed of two frequencies \cite{wigner}, 
wrote down the spherical wave packet in 
the asymptotic region and evaluated the delay in the emergence of the 
wave packet due to interaction. Thus, starting with the asymptotic 
form of a wave packet such as, 
\begin{equation}\label{wavepack}
\Psi (r,t) \simeq 
{1 \over \sqrt{r}} [e^{-i(kr+\omega t)}\, {\rm cos}(r\Delta k + 
t \Delta \omega) \, - \, e^{i(kr-\omega t + 2\delta)} {\rm cos}(r\Delta k - 
t \Delta \omega + 2\Delta \delta)] \, , 
\end{equation}
one finds that the first term has a peak at $r_p = -t (d\omega/dk)$, where 
$d\omega/dk$ is the group velocity. The second term has a peak at 
$r_p = (d\omega/dk) t - 2d\delta/dk$ and the interaction 
has delayed the particle 
by a time $\Delta t = 2 (d\omega/dk)^{-1}\,d\delta/dk = 2 \hbar d\delta/dE$, 
where $\delta$ is the scattering phase shift. 

One could start with a wave packet corresponding to the 
asymptotic form of the full wave function,  
\begin{equation}\label{fullwave}
\Psi_{\vec k}(\vec {r}) = e^{i\vec{k} \cdot \vec{r}} \,+\,f(\vec{k})\, 
{e^{ikr} \over r}\, 
\end{equation}
and choose to perform a partial wave expansion of 
$\Psi_{\vec k}(\vec {r})$, namely, 
\begin{equation}
\Psi_{\vec k}({\vec r}) = 4 \pi \sum_{lm} \,i^l \, {u(l,k,r) \over kr} \, 
Y^*_{lm}(\hat{k})\, Y_{lm}(\hat{r}) \, , 
\end{equation}
such that, 
\begin{equation}\label{radialasymp}
u(l,k,r) = {i \over 2} \, ( e^{-i(kr-l\pi/2)} \,-\,S^l(E) e^{i(kr-l\pi/2)}\,) 
\, ,
\end{equation}
with the $l$ dependent $S$-matrix given by $S^l(E) = \exp(2i \delta_l)$. 
If one now performs the same procedure as that of Wigner, but with 
$2 \delta$ replaced rather by [$\Im$m ln($S^l(E)$)], then one gets an 
$l$-dependent time delay, $\Delta t^l(E)$ given by, 
\begin{equation} \label{radial2}
\Delta t^l(E) = \hbar \, \Im {\rm m} \,{d \over dE} ({\rm ln}S^l(E)) \,
=\,  \Re e [ -i \hbar(\,[S^l]^{\,-1}\, {d S^l / dE}\,)
\,] \, . 
\end{equation}
Depending on the wave packet whose peak we choose to follow, which can either
contain the full asymptotic wave (\ref{fullwave}) or the radial one 
(\ref{radialasymp}),
two different delay times emerge. We perform this analysis for the
multi-channel case.

Equation (\ref{radialasymp}), for scattering with multiple channels takes 
the form,
\begin{equation}
u_{i}(l,k,r) \,= \, \phi_{i}(l,k,r) \,-\, 
\sum_n \,\,S^l_{in}(E) \phi^*_{n}(l,k,r) \, , 
\end{equation}
where $\phi_{i}(l,k,r)$ is the incoming plane wave and $\phi^*_{n}(l,k,r)$ 
are outgoing ones in all possible channels. 
The complex multichannel $S$-matrix elements, $S^l_{in}(E)$, 
can in general be parameterized in terms of an energy dependent function 
$\eta^l_{in}(E)$ and a phase $\xi^l_{in}(E)$ as 
$S^l_{in}(E) \, =\,\eta^l_{in}(E)\, e^{i\xi^{\,l}_{in}(E)}$. 
To evaluate the time delay of a particle entering in the $i^{th}$ channel and 
leaving in the $j^{th}$ one, we re-write 
$2 \xi^l_{ij}\,=\, \Im {\rm m}
\,({\rm ln}\,S^l_{ij})$. Once again, following the peak of the wave packet 
in the radial direction (similar to that 
given by (\ref{wavepack}) but with the incoming and outgoing plane waves 
replaced by the $\phi_{i,j}(l,k,r)$'s mentioned above and 
$2 \delta$ replaced by 
$ \Im {\rm m} \,({\rm ln}\,S^l_{ij})$), one gets, 
\begin{eqnarray}\label{multid}
\Delta t^l_{ij}(E) &=& \hbar {d\over dE}(\Im {\rm m} \,{\rm ln} S^l_{ij})
\\ \nonumber 
&=&\, \Re e [ -i \hbar (\,[S^l_{ij}]^{-1} {d S^l_{ij} / dE}\,)
\,] \, 
\end{eqnarray}
where $S^l_{ij}$ are elements of a multichannel $S$-matrix, 
${\bf S}$, however, for a given 
partial wave $l$. The superscript $l$ was not given by Smith, but was implicit 
in the entire derivation. 
If one does not perform a partial wave expansion, but rather 
prefers to work with the full wave function as in (\ref{fullwave}), then 
the time delay is that of the full wave packet and is proportional to the
argument of the full scattering amplitude $f(\vec{k})$ \cite{nussenrev}. 
For the single channel case, one can simply write, 
\begin{equation}\label{angtimed}
\Delta t^{ang} (E,\theta)\, =\, \hbar {\partial \over \partial E} 
arg [f(k,\theta)]\,.  
\end{equation}
This is an angular time delay of the full wave packet \cite{nussen72} 
and is not trivially 
connected to the time delay of Wigner or Smith even in the single channel 
case. If however, one averages over the angles, this delay becomes 
proportional but not equal 
to the energy derivative of the phase shift as defined by Wigner and 
also obtained by Smith for single channel resonances. In the more general 
case of multichannel resonances, the angular time delay takes the form 
\cite{ohmura}, 
\begin{eqnarray}\label{ohmurtd}
\Delta t_{ij}^{ang} (E,\theta) 
\, &=&\, \hbar {d \over dE} arg (S_{ij} \, - \, 
\delta_{ij}) \\ \nonumber 
&=&  \hbar {d \over dE} [\Im {\rm m \, ln} (S_{ij} \, -\, \delta_{ij})] \, =\, 
\Re e [ -i \hbar (\,[S_{ij} \, - \, \delta_{ij}]^{-1} {d S_{ij} / dE}\,)
\,] \, 
\end{eqnarray}
where, $\delta_{ij}$ is the Kronecker delta and the elements $S_{ij}$ belong 
to the full $S$-matrix which is not decomposed into partial waves. 
Eqs (\ref{multid}) and (\ref{ohmurtd}) appear to be of a similar form when 
$i \ne j$, however, (\ref{multid}) involves an $l$ dependent $S$-matrix and 
(\ref{ohmurtd}) involves the full one which has not been expanded into 
partial waves. We shall see below that 
one cannot directly use the relation (\ref{ohmurtd}) to evaluate the
time delay in a single partial wave $l$.

\section{Observable angular time delay} 
With the time delay of Smith (Eq. \ref{multid}) being given for each value 
of $l$, it is quite useful in analysing hadron resonances which are 
classified according to their $l$ values. 
The angular time delay in (\ref{ohmurtd}), however, corresponds to the 
delay of an entire wave packet. We shall now perform a partial wave expansion 
of the $S$-matrix in order to investigate if this relation can be brought 
into a form useful for characterizing hadron resonances. 
We restrict the discussion to the scattering of spin zero by spin one-half 
particles (since it corresponds to the realistic case of pion-nucleon 
scattering which we will discuss later as an explicit example).
Alternatively, all our formulae are valid for the scattering of spinless
particles.
For a given state of angular momentum $J$, $l$ may take values, $J + 1/2$ 
or $J - 1/2$ and due to conservation of parity, the initial and final state
carry the same angular momentum $l$. 
Neglecting the possibility of a spin flip amplitude for the
present discussion, one can write the partial wave expansion of 
the non-spin-flip amplitude in the case of single-channel scattering as,
\begin{equation}
f(E,\theta)\,=\, \sum_l\,T_l\,P_l({\rm cos} \theta)\, ,
\end{equation}
where the $T$-matrix here is given as, 
$(2l+1) (S_l - 1)/2ik$, with $S_l=\exp (2i\delta_l)$ and $\delta_l$ the 
scattering phase shift.  The argument of the 
scattering amplitude $f(E,\theta)$ 
which appears in the definition of angular time 
delay (\ref{angtimed}), is then,
\begin{equation}
arg f(E,\theta) \,=\,\arctan \,\biggl [ \,{\sum_l |T_l|\, 
P_l({\rm cos}(\theta) \, {\rm sin} (\phi_l) \over 
\,\sum_l |T_l|\, 
P_l({\rm cos}(\theta) \, {\rm cos} (\phi_l)} \,\biggr ]
\end{equation}
with $T_l\,=|T_l|\,e^{i\phi_l}$. The energy derivative of $arg f(E,\theta)$ 
involves an expression entangled in all $l$'s which is not of much use
to analyze hadron resonances. However, if one evaluates an angle averaged 
time delay with the $f$ written in terms of partial waves, one gets,
\begin{equation}
\hbar \,
\int\, |f(E,\theta)|^2 \, {\partial \over \partial E} arg f(E,\theta) \, 
d\Omega \, =\, 4 \pi \,\hbar\, \sum_l \,|T_l|^2\, {d\over dE}(arg\,T_l) \, .
\end{equation}
In the single channel case, $arg \,T_l\,=\,\delta_l$, where $\delta_l$ is the
scattering phase shift, and 
\begin{equation}\label{eisreduce}
\hbar \,
\int\, |f(E,\theta)|^2 \, {\partial \over \partial E} arg f(E,\theta) \, 
d\Omega \, =\, {\pi \over k^2} \, \sum_l \,(2l+1)\,
{\rm 2 \,sin}^2 \delta_l \,\biggl( 2\,\hbar\, {d\delta_l \over dE} \,\biggr )
\, .
\end{equation}
One could now use such a relation for characterizing resonances 
classified by partial waves. 
This angle averaged 
relation is not equal but somewhat similar to that obtained by
Wigner, namely, $2 \,\hbar\,d\delta_l/dE$.

Nussenzveig \cite{nussen72}
introduced an additional term to subtract the time delay in 
the forward direction where the incident beam interferes with the 
scattered one. Performing an angle average as above but with the 
new term included, it is easy to show that for single channel
scattering,
\begin{equation}
\hbar \,
\int\, |f(E,\theta)|^2 \, {\partial \over \partial E} arg f(E,\theta) \, 
d\Omega \, + \,{2\,\pi \over k^2} {d\over dE}[k\,{\rm Re}f(E,0)]\,=\, 
{\pi \over k^2} \, \sum_l \,(2l+1)\,2\,\hbar\, {d\delta_l \over dE} \, .
\end{equation}
One recovers back the Wigner's phase time delay result now. 

\subsection{Eisenbud's unpublished multichannel delay time matrix}
In 1948, in his Ph.D. dissertation \cite{eisen},  
Eisenbud obtained the phase time delay in single channel scattering 
which can also be found 
in \cite{wigner} and is given by the energy derivative of the scattering 
phase shift. He further derived the multichannel generalization by 
restricting to the case where ``all sub-systems have spin zero and 
only waves with $l = 0$ participate in the reaction". He wrote down the 
following relation for a time delay matrix with elements, 
\begin{equation}
\Delta t_{ij}^{Eisen} (E) = {d \over dE} [S\,-\, 1]_{ij} \,.
\end{equation}
This expression is similar to the angular time delay discussed above.
Since Eisenbud did not consider anything but $s$-waves, 
in connection with the example of a constant phase 
multichannel Breit-Wigner resonance, he concluded that his relation 
in the case of single channel scattering, 
reduces to that of the phase time delay defined by him before. 
Eisenbud's time delay for $s$-waves was obviously angle independent. 
However, in the general (and more realistic) case 
of several partial waves participating in a reaction, we have just seen
that (a) it does not reduce to any useful form unless we perform an
angle averaging and (b) even after the angle averaging, it reduces only 
to Eq. (\ref{eisreduce}) which has an extra ${\rm sin}^2 \delta_l$ term as 
compared to the phase time delay.    

\section{Lifetime matrices}
The concept of a lifetime matrix in quantum collisions was first introduced by 
F. T. Smith \cite{smith}. A collision lifetime was first defined as the 
limit, $R \to \infty$, of the difference between the time that the scattering 
particles spend within a distance $R$ of each other, with and without 
interaction. The definition of this time as introduced by Smith, is now 
popularly know as the dwell time and is used in a variety of quantum 
mechanical tunneling problems \cite{dwell}. 
\subsection{Smith's Hermitian matrix}
The collision lifetime was used 
to construct a lifetime matrix ${\bf Q}$, such that ${\bf Q}$ was Hermitian 
and a diagonal element $Q_{ii}$ gave the average lifetime of a collision 
beginning in the $i^{th}$ channel. This ${\bf Q}$ matrix has some elegant 
properties:
\begin{eqnarray}\label{smithslife} 
&&{\bf Q}\, =\, i\, \hbar {\bf S} \,d{\bf S}^{\dagger}/dE\, =\, -i\, \hbar 
(d{\bf S}/dE)\, {\bf S}^{\dagger} \, =\, {\bf Q}^{\dagger}  \\ \nonumber
&&{\bf Q}\, =\,-\, {\bf S} \, {\bf t}\, {\bf S}^{\dagger} \, =\, 
({\bf t}\, {\bf S})\, {\bf S}^{\dagger} \\ \nonumber
&&{\bf S}\, =\,{\bf 1} \, -\, (i/\hbar)\, \int_E^{\infty}\, {\bf Q}(E')\, 
{\bf S}(E')\, dE'\, .
\end{eqnarray}
The first line of these equations shows that ${\bf Q}$ is Hermitian. 
The second line identifies a time operator and the third one shows that one can 
find ${\bf S}$ from ${\bf Q}$ as a function of energy, by inverting the 
relation given above. The diagonal elements $Q_{ii}$ define an average delay 
time for a particle injected in the $i^{th}$ channel. Since the particle 
has probability $|S_{ij}|^2$ of emerging in the $j^{th}$ channel, one can 
write,
\begin{eqnarray}
\langle \Delta t^l_{ii} \rangle (E) \, &=&\,\sum_j \,S^{*\,l}_{ij}
\, S^l_{ij} \, 
\Delta t^l_{ij} \\ \nonumber
&=&\, \Re e [ -i \hbar \, \sum_j \, S^{* \,l}_{ij} \, 
dS^l_{ij}/dE \,] \, =\, Q^l_{ii}\,.
\end{eqnarray}
The angular brackets indicate the average over channels. 
We remind the reader once again that the above $S$-matrix elements, 
$S^l_{ij}$ and hence the $Q$-matrix elements, $Q^l_{ii}$ (or $Q_{ii}$ as in 
Smith's paper), are for a given value of $l$. 

\subsection{Ohmura's angle dependent lifetime matrix} 
One can try to construct an angle dependent 
lifetime matrix from the angular time delay
expression (\ref{ohmurtd}) in the same spirit and one 
obtains \cite{ohmura},
\begin{eqnarray}
\langle \Delta t^{ang}_{ii} \rangle (E, \theta) \, &=&\,\sum_j 
\,(S^*_{ij} \,-\, \delta_{ij})\, (S_{ij} \,-\, \delta_{ij}) \, 
\Delta t^{ang}_{ij} \\ \nonumber  
&=&\, \Re e [ -i \hbar \, \sum_j \, (S^*_{ij}\, -\, \delta_{ij}) 
\, dS_{ij}/dE \,] \, =\, Q^{ang}_{ii}(E, \theta)\,.
\end{eqnarray}
Note that the average time delay in the angular case involves weighting the 
$\Delta t^{ang}_{ij}$ ($=    
\Re e [ -i \hbar (\,[S_{ij} \, - \, \delta_{ij}]^{-1} {d S_{ij} / dE}\,)
\,] \,
$) by the probabilities, $|S_{ij} \,-\, \delta_{ij}|^2$, 
with $S_{ij}$ being elements of the full $S$-matrix with no partial wave 
expansion. It can be easily seen that,
\begin{equation}
{\bf Q}^{ang}(E, \theta) \,
=\,i \,\hbar\, ({\bf S} \,-\, {\bf 1})\, {d \over dE} 
({\bf S} \,-\, {\bf 1})^{\dagger} \, \ne\, 
-\,i \,\hbar\, \biggl [\,{d \over dE} ({\bf S} \,-\, {\bf 1}) \,\biggr]
\, ({\bf S} \,-\, {\bf 1})^{\dagger} \,=\, 
{\bf Q}^{ang}\, ^{\dagger}(E, \theta)\, .
\end{equation}
The elements of ${\bf Q}^{ang}$, namely, 
\begin{equation}\label{ohmurslife}
Q^{ang}_{ij}(E, \theta) \, =\, i\, \hbar \, \sum_n \, (S_{in} \, - \, \delta_{in}) \, 
dS^*_{jn}/dE \, ,
\end{equation}
are not symmetric ($Q^{ang}_{ij} \ne Q^{ang}_{ji}$) and hence the time delay
is not time reversal invariant. Indeed, it would be odd to expect of a 
time delay which depends on the angle of the outgoing particles to be time 
reversal invariant. 
 
\subsection{Angle and channel averaged time delay}
In the previous section, we saw that in the single channel scattering, if one
averages the angular time delay over all directions and subtracts a 
forward scattering term, one indeed recovers Wigner's phase time delay which
is consistent with Smith's time delay matrix expression (\ref{multid}). 
We now demonstrate that using a similar procedure one can reduce the 
average time delay (average over channels) given by Ohmura to the 
$l$ dependent average time delay, $Q^l_{ii}$, given by Smith.
For a particle entering in the $i^{th}$ channel and with the possibility of 
emerging in 
some $j^{th}$ channel, we begin by considering the angle and channel 
averaged expression, namely, 
\begin{equation}\label{nussgeneral}
\langle \Delta t_{ii}^{ang} \rangle_{av} (E)\,=\, 
\hbar \,\sum_j \, 
\int\, |f_{ij}(E,\theta)|^2 \, {\partial \over \partial E} 
arg f_{ij}(E,\theta) \, 
d\Omega \, + \,{2\,\pi \, \hbar \over k^2} {d\over dE}[k\,\delta_{ij} \,
\Re {\rm e}f_{ij}(E,0)]\,. 
\end{equation}
The angular brackets indicate the average over channels, and the subscript 
$av$ indicates the average over directions. 
Expressing the $f_{ij}$ in terms of partial waves as before, 
noticing that $T^l_{ij} = (S^l_{ij} \,-\, \delta_{ij})/2ik$ and using 
as before, $arg \,(S^l_{ij} \,-\, \delta_{ij}) \,=\, \Im {\rm m}\,\,  
{\rm ln} \,[S^l_{ij} \, -\,\delta_{ij}]$, 
the first term on the right hand side of the above equation becomes, 
\begin{eqnarray}
\langle \Delta t_{ii}^{ang} \rangle_{av\,1} &=& 4 \pi \,\hbar \sum_j \, 
\sum_l \,(2l+1)\, |T^l_{ij}|^2\, {d\over dE}(arg \,T^l_{ij}) \\ \nonumber 
&=& {\pi \over k^2} \, \sum_l \, (2l+1)\,\sum_j \, (S^l_{ij}\, -\, 
\delta_{ij})^*\,(S^l_{ij}\, -\, \delta_{ij})\,\Re e \biggl [\,-i \,\hbar
\,(S^l_{ij} \,-\,\delta_{ij})^{-1} \,{d \over dE}S^l_{ij} \, \biggr ] \\ 
\nonumber
&=& {\pi \over k^2} \, \sum_l \, (2l+1)\,\sum_j \, (S^l_{ij}\, -\, 
\delta_{ij})^*\,(S^l_{ij}\, -\, \delta_{ij}) \, [\Delta t^l_{ij}]^{new}\, .
\end{eqnarray}
Starting with the angle dependent multichannel time delay matrix, 
we have now obtained a new definition of an angle averaged and $l$ dependent 
time delay matrix with elements $[\Delta t^l_{ij}]^{new}$. 
Since the factor $(S^l_{ij}\, -\, 
\delta_{ij})^*\,(S^l_{ij}\, -\, \delta_{ij})$ is real, one can re-write the 
above equation as,
\begin{eqnarray}\label{nussgeneral1}
\langle \Delta t_{ii}^{ang} \rangle_{av \,1} &=& {\pi \over k^2} \, 
\sum_l \, (2l+1) \, \Re e \biggl [\, - i \, \hbar \, 
\sum_j \, (S^l_{ij}\, -\, \delta_{ij})^* \,
{d \over dE}S^l_{ij}  \biggr ] \\ \nonumber
&=& {\pi \over k^2} \,\sum_l \, (2l+1) \, \Re e \biggl [\, \biggl (\,
- i\, \hbar\,\sum_j\, S^{*\,l}_{ij}\, \,{d \over dE}S^l_{ij} \biggr ) \, +\,
i\,\hbar\, {d\over dE}S^l_{ii}\, \biggr]\\ \nonumber
&=& {\pi \over k^2} \,\sum_l \, (2l+1)\, \biggl [\,Q^l_{ii}\, -\,\hbar\, 
\Im {\rm m}\,{d\over dE}S^l_{ii}\, \biggr ]\, .
\end{eqnarray}
The $Q^l_{ii}$ above is indeed a diagonal element of Smith's lifetime 
matrix and hence also his channel averaged time delay for a particle 
being injected in the channel $i$. If one now evaluates the second term in 
(\ref{nussgeneral}) in a similar way as the first, it is easily verified
to exactly cancel the second term in the last line of 
(\ref{nussgeneral1}), such that, 
\begin{equation}
\langle \Delta t_{ii}^{ang} \rangle_{av} (E)\, =\, {\pi \over k^2} \,\sum_l \, (2l+1)
\,Q^l_{ii}(E)\, .
\end{equation}
 
\section{Negative time delay}
The relevance of a ``delay time" for resonances comes from the fact that 
the creation and propagation of an unstable intermediate state in a 
scattering reaction, delays the reaction. Below, we shall discuss 
different aspects of negative time delay and try to clarify some 
misconceptions in literature.

\subsection{Misinterpreted misnomer}
In a recent work \cite{habwork}, 
the example with a Breit-Wigner resonance from Eisenbud's
unpublished thesis was reproduced to demonstrate that the time delay 
relations due to Smith and Eisenbud differ. 
The authors, however, wrongly interpreted a misnomer to be an error which
propagated in literature due to Smith. 
It is probably due to the fact that Eisenbud's 
thesis remained unpublished that the multichannel 
time delay relation (\ref{multid}) which is 
entirely due to Smith was often mentioned to be due to Eisenbud. 
Repeating the example from Eisenbud's thesis 
of a two channel Breit-Wigner resonance with one phase in the $T$-matrix, 
the authors reached the conclusion 
that Eisenbud's time delay is positive and the same for both channels in 
contrast to that due to Smith which can be negative for a channel with
a branching fraction less than half. There are several points worth noting:\\
(i) The general form of Eisenbud's time delay which depends on the $S$-matrix 
which has not been decomposed into partial waves  
cannot be directly compared with that of Smith which is written for a 
particular $l$ (see Eqs (\ref{multid} - \ref{ohmurtd}) and the discussion
below Eq. (\ref{ohmurtd})). \\
(ii) If one assumes the existence of only $s$-waves in reactions (as Eisenbud 
did in his thesis), then the two delay times can be compared. However, as we
shall see below, even in this case, there is no need for Eisenbud's time delay
to be equal in both channels and positive.\\
(iii) Probably not 
having noticed that the lifetime matrix definition for an angular 
time delay differs from that of the radial time delay as given by Smith
(see Eqs (\ref{smithslife}) to (\ref{ohmurslife})), on the one hand 
the authors in \cite{habwork} claimed that Smith's work has propagated an error 
(which was actually a misnomer) 
in literature and on the other hand used his definition of the lifetime 
matrix (which is consistent with his time delay matrix). 
This definition was indeed used to demonstrate the validity of speed plots
for characterizing resonances.\\
(iv) Referring to Smith's radial time delay in the multichannel case as 
a misquoted relation, it was inferred that the conclusions regarding negative 
time delay in \cite{me1,usall} were wrong. In what follows, we shall see how 
regions of negative time delay manifest themselves with the opening 
of inelastic channels in elastic scatterings. 

In order to be able to compare the angular and radial time delay relations, 
let us restrict ourselves 
to the situation in Eisenbud's thesis where only $s$-waves 
participate in reactions. 
We start with a multichannel, unitary,
$ 2 \times 2 $ matrix \cite{goebel}, 
\begin{equation}\label{4}
{\bf S} \,=\,\left(
\begin{array}{cc}
\eta \,e^{2i\delta _1} & i\,(1 - \eta^2)^{1/2}\,e^{i\,(\delta _1 + \delta _2)
}\\
i\,(1 - \eta ^2)^{1/2}\,e^{i\,(\delta _1 + \delta _2)} & \eta \, e^{2i\delta _2}
\end{array}
\right )
\end{equation}
where $\delta _1$ and $\delta _2$ are the real scattering phase shifts for
the elastic scattering ($1 \rightarrow 1$ and $2 \rightarrow 2$) in channels
$1$ and $2$ and $\eta$ is the inelasticity parameter
(with $0 < \eta \le 1$). Smith's radial time delay matrix then takes the form,
\begin{equation}\label{5}
\Delta {\bf t}^{Smith} \,=\,
2 \hbar \, \, \left(
\begin{array}{cc}
{d\delta _1 / dE} \,\,\, &\,\,\, 
{1 \over 2}{d \over dE} (\delta _1 + \delta _2)\\
\\
{1 \over 2}{d \over dE}(\delta _1 + \delta _2)\,\,\, & \,\,\,{d\delta _2 / dE}
\end{array}
\right )
\end{equation}
The time delay matrix of Eisenbud has elements with terms involving energy
derivatives of the phase shift as well as inelasticity. For example,
\begin{equation}\label{eistee}
\Delta t_{ii}^{Eisenbud} \, =\, {2\,[\eta^2 - \eta {\rm Cos} (2 \delta_i) ] 
{d\delta_i / dE} \, - \, {\rm Sin} (2\delta_i) {d\eta / dE} \over 
1\, +\, \eta^2\, -\, 2\eta {\rm Cos} (2\delta_i)}\, . 
\end{equation}
It can be seen that $\Delta t_{11}^{Eisenbud}$
may not necessarily be the same as $\Delta t_{22}^{Eisenbud}$ for the obvious
reason that $\delta_1$ may not be the same as $\delta_2$. Both the delay 
times, due to Smith and Eisenbud may not necessarily be positive in the 
two channels. As an example, we consider the phase shifts and the $t$-matrix 
phase for pion-nucleon elastic scattering 
in the vicinity of the nucleon resonances, N$^*$(1510) and N$^*$(1655), 
where the masses in parentheses correspond to the pole positions. 
These resonances occur in the $l=0$ partial wave, 
with total spin $J=1/2$ and isospin $I=1/2$ 
(with the label $S_{11}$ in the notation $l_{2J,2I}$).  
\begin{figure}[ht]
\includegraphics[width=9cm,height=12cm]{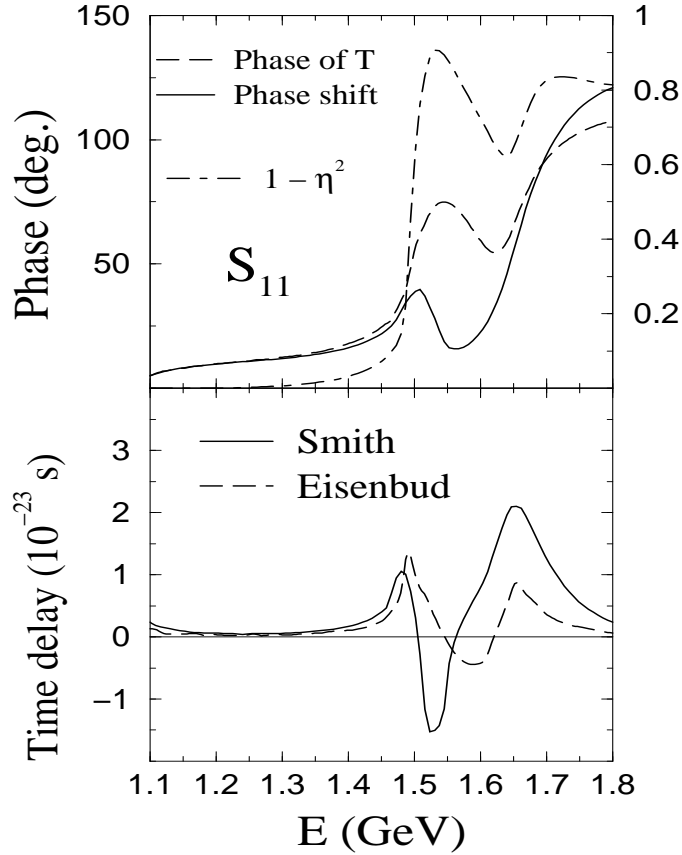}
\caption{\label{fig:eps1}
The $T$-matrix phase (dashed lines) and scattering phase shifts (solid lines)
for the $S_{11}$ resonances occurring in pion-nucleon elastic scattering. 
The dot-dashed lines are
inelasticities (1 - $\eta^2$) with the scale given on the right. The lower
part of the figure contains the time delay plots using Smith's and
Eisenbud's relations.}
\end{figure}
One can see that the regions of negative time delay grow with increasing 
inelasticity. Both the N$^*$ resonances nevertheless show up as positive 
peaks in both the Eisenbud and Smith delay times. 
\subsection{Density of states and the classical example of averages} 
Beth and Uhlenbeck \cite{BU}, while calculating
virial coefficients $B, C$
in the equation of an ideal gas: $pV=RT\left[1 +\frac{B}{V} + 
\frac{C}{V^2} + \cdots \right]$, found
that the difference between the density of states
(of scattered particles) with interaction $dn_l(E)/dE$
and without $dn^{(0)}_l(E)/dE$ is,
\begin{equation} \label{BU1}
\frac{dn}{dE}=\frac{dn_l(E)}{dE} -\frac{dn^{(0)}_l(E)}{dE} =\frac{2l+1}{\pi}
\frac{d\delta_l(E)}{dE}\, .
\end{equation}
In a resonant scattering, this is the density of states
of a resonance (in terms of the decay products) \cite{nonexpopapers}. 
For instance
$T=(\Gamma_R/2)/(E_R -E -i\Gamma_R/2)$, gives,
\begin{equation} \label{BU2} 
\frac{d\delta}{dE}=\frac{\Gamma_R/2}{(E_R-E)^2 +\Gamma_R^2/4} \, .
\end{equation}
This interpretation is also useful in analysing the survival probabilities
of resonances in all partial waves and at all energies 
\cite{nonexpopapers} except the near threshold 
$s$-wave resonances \cite{meprl}. 
The right hand side of the Beth-Uhlenbeck formula can be seen to be the 
so-called phase time delay (and so are the diagonal elements of 
Smith's multichannel time delay matrix (\ref{5})).
In the case of $s$-wave resonances, 
the phase and dwell time delay become equal at high energies \cite{meprl}. 
Coming back to Smith's definition of 
time delay which is the difference between the time with
and without interaction, taken together with the results in \cite{iangaspa} 
which showed that dwell times correspond to densities of states in a 
region, the connection between time delay and the difference in the 
density of states becomes more transparent.  

In the case of single channel scattering, 
one expects the time delay of Wigner and 
Smith to peak in the vicinity of a resonance. This is intuitively
based on the fact that an unstable particle (seen as an intermediate state
becoming on-shell) has a finite lifetime which delays the reaction.
In such a case, the time interval $\delta t$, 
spent by the particle during the scattering process is 
$\delta t\, =\,\delta t_0 \,+ \,\Delta t$ 
where $\Delta t > 0$ is the contribution due to interaction
and $\delta t_0$ is the time spent without interaction.
In the case of multi-channel resonances, this picture has to be
modified. If the branching ratio, $\Gamma_i/\Gamma$ ($\Gamma_i$ and
$\Gamma$ are the partial and full widths, respectively) of the 
elastic channel $i$ is close to one, the validity of the above
interpretation remains. Smaller branching ratios lead, however, to negative
delay times ($\Delta t < 0$) due to the loss of flux in this channel. We
can explain it by using classical averages which mimic the quantum 
ones in the delay times. 
In the multi-channel case, $\delta t$ is now the time 
$\delta t_{\rm interaction}$ say, 
weighted with the probability of arrival. 
If we start with $N$ particles scattering on a target, due to absorption (i.e.
the opening of a new channel) only $N_a < N $ will emerge after scattering.
Hence, $\delta t\,=\,(N_a/N)\,\delta t_{\rm interaction}$, but 
$\delta t_0$ remains unchanged. As a result, we obtain
$\Delta t\,=\,(N_a/N)\,\delta t_{\rm interaction}-\delta t_0$.
If due to a resonance $\delta t_{\rm interaction} > \delta t_0$,
it can still happen that $(N_a/N)\delta t_{\rm interaction} < 
\delta t_0$, and therefore $\Delta t < 0$ if $N_a/N$ is small
corresponding to a small branching ratio in this channel.
This classical picture is not very far from what really happens
to the quantum concept of time delay. Indeed, in
Figure 1, $\Delta t$ has a negative dip exactly at the very same 
position where the inelasticity is largest.

\section{Conclusions}
According to the wave packet analysis of the scattered wave, there
are only two viable time delay concepts in quantum
scattering. The first one is due to Smith involving 
the $S$ matrix elements after a partial wave decomposition has been
performed. As a result, this time delay depends only on energy.
The other possibility is conceptually similar, but makes use of the
full $T$-matrix ($S\,=\,1\,+\,2i\,T$) and is energy and angle dependent.
Whereas Smith's approach is more suitable for resonant scattering with
resonances classified  according to the $l$ quantum number, the
angular time delay is a directly measurable concept.
Not even in a single channel scattering formalism will these two agree.
Only in an artificial world of a single channel with the restriction to 
$s$-waves only, we obtain $arg \,f \,=\,arg \,T_{l=0}\,=\,arg \,S_{l=0}$ 
as considered by Eisenbud who introduced the time delay concept. 
It is however possible to make a connection between the two quantum time 
concepts. Averaging the angular time delay over angles and channels
gives us back the {\it average} time delay of Smith.


\begin{thebibliography}{99}
\bibitem{eisen}
L. E. Eisenbud, Ph.D. Thesis, Princeton University, unpublished (1948).
\bibitem{wigner}
E. P Wigner, Phys. Rev. {\bf 98}, 145 (1955). 
\bibitem{others}
D. Bohm, Quantum Theory, Prentice-Hall, Englewood Cliffs, NJ, p. 257 (1951); 
A. M. Lane and R. G. Thomas, Rev. Mod. Phys. {\bf 30}, 257 (1958).
\bibitem{goldb}
M. Froissart, M. L. Goldberger and K. M. Watson, Phys. Rev. {\bf 131}, 
2820 (1963); M. L. Goldberger and K. M. Watson, Collision Theory, 
Wiley, New York (1964).
\bibitem{smith}
F. T. Smith, Phys. Rev. {\bf 118}, 349 (1960).
\bibitem{chaos}
Yan V. Fyodorov and Hans-J\"urgen Sommers, Phys. Rev. Lett. {\bf 76}, 
4709 (1996);
Dmitry V. Savin, Yan V. Fyodorov and Hans-Juergen Sommers, 
Phys. Rev. E {\bf 63}, 035202 (2001); 
R. O. Vallejos, A.M. Ozorio de Almeida and C.H. Lewenkopf, 
J. Phys. A {\bf 31}, 4885 (1998).
\bibitem{haugewin}
H. Winful, Phys. Rep. {\bf 436}, 1 (2006); 
E. H. Hauge and J. A. St\o vneng, Rev. Mod. Phys. {\bf 61}, 917 (1989);
E. H. Hauge, J. P. Falck and T. A. Fjeldly, Phys. Rev. {\bf B 36},
4203 (1987). 
\bibitem{lipp}
B. A. Lippmann, Phys. Rev. {\bf 151}, 1023 (1966). 
\bibitem{martin}
Ph. A. Martin, Acta Phys. Austrica Suppl. {\bf 23}, 157 (1981).
\bibitem{atomic}
G. Isi\'c {\it et al}., Phys. Rev. A {\bf 77}, 033821 (2008); 
W. O. Amrein and Ph. Jacquet, Phys. Rev. A {\bf 75}, 
022106 (2007); N. Yamanaka, Y. Kino, and A. Ichimura, 
Phys. Rev. A {\bf 70}, 062701 (2004); Chun-Woo Lee, Phys. Rev. A {\bf 58}, 
4581 (1998).   
\bibitem{me1}
N. G. Kelkar, J. Phys. G: Nucl. Part. Phys. {\bf 29}, L1 (2003).
\bibitem{usall}
N. G. Kelkar, M. Nowakowski, K. P. Khemchandani
and S. R. Jain, Nucl. Phys. {\bf A730}, 121 (2004);
N. G. Kelkar, M. Nowakowski and K. P. Khemchandani,
Mod. Phys. Lett. {\bf A 19}, 2001 (2004);
{\it ibid}, Nucl. Phys. {\bf A 724}, 357 (2003); {\it ibid},
J. Phys. {\bf G 29}, 1001 (2003).
\bibitem{nussenrev}
C. A. A. de Carvalhoa and H. M. Nussenzveig, Phys. Rep. {\bf 364}, 83 (2002).
\bibitem{nussen72}
H. M. Nussenzveig, Phys. Rev. D {\bf 6}, 1534 (1972).  
\bibitem{ohmura}
T. Ohmura, Suppl. of the Prog. Theor. Phys. {\bf 29}, 108 (1964).
Ohmura omits the notation where explicit angle dependence is manifest.
However, such a dependence is obvious unless
one replaces 
$arg(S-1)$ by $arg (S^l-1)$ which, however, is not the result
from the scattered wave packet analysis. 
\bibitem{dwell}
Y. Wang, N. Zhu, J. Wang and H. Guo, Phys. Rev. B {\bf 53}, 16408 (1996);
V. S. Olkhovsky and E. Recami, Phys. Rep. {\bf 214}, 339 (1992); 
H. G. Winful, Phys. Rev. Lett. {\bf 91}, 260401 (2003).
\bibitem{habwork}
H. Haberzettl and R. Workman, Phys. Rev. C {\bf 76}, 058201 (2007). 
\bibitem{goebel}
C. J. Goebel and K. W. McVoy, Phys. Rev. {\bf 164}, 1932 (1967). 
\bibitem{BU}
E. Beth and G. E. Uhlenbeck, Physics {\bf 4}, 915 (1937);
K. Huang, {\it Statistical Mechanics} (Wiley, New York, 1963).
\bibitem{nonexpopapers}
N. G. Kelkar, M. Nowakowski and K. P. Khemchandani, Phys. Rev. C {\bf 70}, 
024601 (2004); M. Nowakowski and N. G. Kelkar, 
``Long tail of quantum decay from scattering data", 
to appear as AIP Proceedings of the
``SCADRON70 International Workshop on
Scalar Mesons and Related Topics'' held in Lisbon, Portugal in February 2008.
\bibitem{meprl}
N. G. Kelkar, Phys. Rev. Lett. {\bf 99}, 210403 (2007). 
\bibitem{iangaspa}
G. Iannaccone, Phys. Rev. B {\bf 51}, R4727 (1995); 
V. Gasparian and M. Pollak, Phys. Rev. B {\bf 47}, 2038 (1993). 
\end{thebibliography}
\end{document}